\documentclass[aps,twocolumn,showpacs]{revtex4}
\usepackage{amsmath}
\usepackage{epsfig}

\begin{document}

\title{$P_{c}$-like pentaquarks in hidden strange sector}
\author{Hongxia Huang$^1$, Jialun Ping$^1$\footnote{Corresponding
 author: jlping@njnu.edu.cn}}

\affiliation{$^1$Department of Physics and Jiangsu Key Laboratory for Numerical
Simulation of Large Scale Complex Systems, Nanjing Normal University, Nanjing 210023, P. R. China}

\begin{abstract}
Analogous to the work of hidden charm molecular pentaquarks, we study possible hidden strange molecular pentaquarks composed of $\Sigma$ (or $\Sigma^{*}$) and $K$ (or $K^{*}$) in the framework of quark delocalization color screening model. Our results suggest that the $\Sigma K$, $\Sigma K^{*}$ and $\Sigma^{*} K^{*}$ with $IJ^{P}=\frac{1}{2}\frac{1}{2}^{-}$ and $\Sigma K^{*}$, $\Sigma^{*} K$ and $\Sigma^{*} K^{*}$ with $IJ^{P}=\frac{1}{2}\frac{3}{2}^{-}$ are all resonance states by coupling the open channels. The molecular pentaquark $\Sigma^{*} K$ with quantum numbers $IJ^{P}=\frac{1}{2}\frac{3}{2}^{-}$ can be seen as a strange partner of the LHCb $P_{c}(4380)$ state, and it can be identified as the nucleon resonance $N^{*}(1875)$ listed in PDG. The $\Sigma K^{*}$ with quantum numbers $IJ^{P}=\frac{1}{2}\frac{3}{2}^{-}$ can be identified as the $N^{*}(2100)$, which was experimentally observed in the $\phi$ photo-production.
\end{abstract}


\maketitle

\setcounter{totalnumber}{5}

\section{\label{sec:introduction}Introduction}

The multi-quark study is essential for understanding the low energy quantum chromodynamics (QCD), because
the multi-quark states can provide information unavailable for $q\bar{q}$ meson and $q^3$ baryon, especially
the property of hidden color structure. The pentaquark is one of the important topics of the multi-quark study.
In 2015, the observations of two hidden-charm pentaquarks $P_{c}(4380)$ and $P_{c}(4450)$ at LHCb~\cite{LHCb}
invoked a renewed interest in the pentaquark states. The JLab also proposed to search for these two
$P_{c}$ states by using photo-production of $J/\psi$ at threshold~\cite{Jlab}. Various interpretations of the
hidden-charm pentaquarks have been discussed and many other possible pentaquarks were also proposed in the literatures~\cite{ChenHX,ChenR,HeJ1,ChenHX2,Roca,Oller,Kubarovsky,Guo,Liuxh,Burns,HuangHX1}.

Analogous to the hidden-charm pentaquarks $P_{c}$ states, one may consider the existence of possible $P_{c}$-like
pentaquarks in hidden strange sector, in which the $c\bar{c}$ is replaced by the $s\bar{s}$. In fact, as early as
2001, a $\phi-N$ bound state was proposed by Gao {\em et al.}~\cite{Gao}, which is an analogy to the work of
Refs.~\cite{Brodsky,Luke}, in which they suggested that the QCD van der Waals interaction, mediated by multi-gluon
exchanges, will dominate the interaction between two hadrons when they have no common quarks and this supported the
prediction of nucleon-charmonium bound state near the charm production threshold.
In addition, Liska {\em et al.}~\cite{Liska} demonstrated the feasibility to search for the $\phi-N$ bound state
from $\phi$ meson subthreshold production; some chiral quark model calculation~\cite{Chiral} and lattice QCD
calculation~\cite{Lattic} also support the existence of such a kind of bound state. Very recently, Xie and Guo
studied the possible $\phi p$ resonance in the $\Lambda_{c}^{+} \rightarrow \pi^{0}\phi p$ decay by considering
a triangle singularity mechanism~\cite{Xie}. Our group also investigated the $\phi-N$ bound state in the quark delocalization
color screening model (QDCSM)~\cite{Gao2}, performed a Monte Carlo simulation of the bound state production with
an electron beam and a gold target, and found it was feasible to experimentally search for the $\phi-N$ bound state
through the near threshold $\phi$ meson production from heavy nuclei. In Ref. ~\cite{Gao2}, we only focus on the
$\phi-N$ bound state, however, we also found that the interaction between $\Sigma$ (or $\Sigma^{*}$) and $K$
(or $K^{*}$) was strong enough to form bound states, which is similar to that of $\Sigma_{c}$ (or $\Sigma_{c}^{*}$)
and $D$ (or $D^{*}$)~\cite{HuangHX1}. Since the $P_{c}(4380)$ and $P_{c}(4450)$ are close to the thresholds of the
$\Sigma_{c}^{*} D$ and $\Sigma_{c}D^{*}$, many work studied two $P_{c}$ states as the molecular states composed of
$\Sigma_{c}$ (or $\Sigma_{c}^{*}$) and $D$ (or $D^{*}$) ~\cite{ChenR,HeJ1}. Therefore, we expect the existence of
some molecular states consisted of $\Sigma$ (or $\Sigma^{*}$) and $K$ (or $K^{*}$), which are analogous to the
$P_{c}$ state.

In fact, the pentaquarks composed of light quarks has a very long history. The $\Lambda(1405)$ resonance was
explained as an $N\bar{K}$ molecular state since the 1960s~\cite{Dalitz,Kaiser,Oset,Krippa,Oller2,Jido,Hall}.
The quantities of nucleon resonances near 2 GeV were still unclear both in theory and experiment. Some nucleon resonances
were investigated by coupling with pentaquark channels. One peculiar state is the $N^{*}(1535)$ resonance with spin parity $J^{P}=1/2^{-}$, which is found to couple strongly to the pentaquark channels with strangeness~\cite{Inoue,LiuBC,Helminen,ZouBS1,AnCS,ZouBS2}. Another $J^{P}=1/2^{-}$ nucleon resonance is the
$N^{*}(1895)$, which is a two-star state in the compilation of Particle Data Group (PDG)~\cite{PDG}. However,
its existence is supported by the analysis of the new $\eta$ photo production data~\cite{Kashevarov,Collins},
which showed that the $N^{*}(1895)$ is crucial to describe the cusp observed in the $\eta$ photo production around
$1896$ MeV. Moreover, Refs.~\cite{Kashevarov,Collins} suggested that this $N^{*}(1895)$ had strong coupling to the
$N \eta$ and $N \eta^{\prime}$ channels. In our previous work, we found a $J^{P}=1/2^{-}$ bound state with a mass
varying from $1873$ to $1881$ MeV, and the main component is $N \eta^{\prime}$~\cite{Gao2}, which could correspond to the resonance
$N^{*}(1895)$. Four $J^{P}=3/2^{-}$ nucleon resonances, $N^{*}(1520)$, $N^{*}(1700)$, $N^{*}(1875)$, and $N^{*}(2120)$
are listed in new versions of the PDG~\cite{PDG}, among which a three-star $N^{*}(1875)$ and a two-star $N^{*}(2120)$
still have various interpretations about their internal structures~\cite{Anisovich,HeJ2,HeJ3}. J. He investigated both $N^{*}(1875)$ and $N^{*}(2120)$. He interpreted the $N^{*}(1875)$ as a hadronic molecular state from the $\Sigma^{*} K$ interaction~\cite{HeJ3}, and showed that the $N^{*}(2120)$ in the $K \Lambda(1520)$ photo-production was assigned as a naive three-quark state in the constituent quark model~\cite{HeJ2,HeJ4}.
Besides, the structure near $2.1$ GeV in the $\phi$ photo-production showed
an enhancement in the same energy region as that of $N^{*}(2120)$~\cite{Mibe,Kiswandhi1,Kiswandhi2}. A recent analysis
suggested that it has a mass of $2.08 \pm 0.04$ GeV and quantum number of $J^{P}=3/2^{-}$~\cite{Chang1,Chang2,Qian,Kiswandhi3}. Ref.~\cite{HeJ5} denoted this state as $N^{*}(2100)$, and investigated it from the $\Sigma K^{*}$ interaction on the
hadron level in a quasipotential Bethe-Saltpeter equation approach. So it is also interesting to study the $\Sigma$
(or $\Sigma^{*}$) and $K$ (or $K^{*}$) interactions on the quark level to investigate the possibility of interpreting
these nucleon resonances as hadronic molecular states.

Generally, one of the important ways to generate and identify multi-quark states is the hadron-hadron scattering process.
The multi-quark state will appear as a resonance state in the scattering process.
Therefore, to provide the necessary information for experiment to search for the multi-quark states,
we should not only calculate the mass spectrum but also study the corresponding scattering process.
By using the constituent quark models and the resonating group method (RGM)~\cite{RGM}, we have obtained the $d^{*}$ resonance
in the $NN$ scattering process, and we found that the energy and the partial decay width to the $D$-wave of
$NN$ are consistent with the experiment data~\cite{Ping2009}. Extending to the pentaquark system, we investigated the $N\phi$ state
in the different scattering channels: $N\eta^{\prime}$, $\Lambda K$, and $\Sigma K$~\cite{Gao}. Both the resonance
mass and decay width were obtained, which provided the necessary information for experimental searching at
JLab. Therefore, it is interesting to extend such study to the molecular states composed of $\Sigma$ (or $\Sigma^{*}$) and $K$ (or $K^{*}$). In this work, we will investigate the scattering process of the corresponding open channels to search for any possible resonance states composed of $\Sigma$ (or $\Sigma^{*}$) and $K$ (or $K^{*}$).

It is a general consensus that quantum chromodynamics (QCD) is the fundamental theory of the strong interaction in the perturbative region. However, it is difficult to use QCD directly to study complicated systems in the low-energy region. The QCD-inspired models, incorporating the properties of low-energy QCD: color confinement and chiral symmetry breaking, are still powerful tools to obtain physical insights into many phenomena of the hadronic world. Among these phenomenological models, the quark
delocalization color screening model (QDCSM), which was developed in the 1990s with the aim of explaining the similarities between nuclear (hadronic clusters
of quarks) and molecular forces~\cite{QDCSM0}, has been quite successful in reproducing the energies of the baryon ground states, the properties of deuteron, the nucleon-nucleon ($NN$) and the hyperon-nucleon ($YN$) interactions~\cite{QDCSM1}. In this model, quarks confined in
one cluster are allowed to delocalize to a nearby cluster, and the delocalization parameter is determined by
the dynamics of the interacting quark system, which allows the quark
system to choose the most favorable configuration through its own
dynamics in a larger Hilbert space. Besides, the confinement interaction between quarks in different cluster orbits is
modified to include a color screening factor, which is a model description of the hidden color channel coupling
effect~\cite{HuangHX2}. Recently, this model has been used to study the hidden-charm pentaquarks~\cite{HuangHX1}. We found that the interaction between $\Sigma_{c}$ (or $\Sigma_{c}^{*}$) and $D$ (or $D^{*}$) was strong enough to form some bound states, and $P_{c}(4380)$ can be interpreted as the molecular state $\Sigma^{*}_{c}D$ with quantum numbers $IJ^{P}=\frac{1}{2}\frac{3}{2}^{-}$.

In this work, we study the molecular states of $\Sigma$ (or $\Sigma^{*}$) and $K$ (or $K^{*}$), calculate both the mass and decay widths of these states, analyze the possibility of the $P_{c}$-like pentaquarks in hidden strange sector, and interpret some nucleon resonances as hadronic molecular states. In the next section, the framework of the QDCSM is briefly
introduced. Section III devotes to the
numerical results and discussions. The summary is shown in the
last section.

\section{The quark delocalization color screening
model (QDCSM)}

The quark delocalization color screening
model has been widely described in the literatures~\cite{QDCSM0,QDCSM1}, and we refer the reader to those works for details. Here, we
just present the salient features of the model. The model
Hamiltonian is:
\begin{widetext}
\begin{eqnarray}
H & = & \sum_{i=1}^5\left(m_i+\frac{p_i^2}{2m_i}\right)-T_{CM} +\sum_{j>i=1}^5
\left(V^{C}_{ij}+V^{G}_{ij}+V^{\chi}_{ij} \right), \\
V^{C}_{ij} & = & -a_{c} \boldsymbol{\lambda}^c_{i}\cdot \boldsymbol{
\lambda}^c_{j} ({r^2_{ij}}+v_{0}), \label{sala-vc} \\
V^{G}_{ij} & = & \frac{1}{4}\alpha_s \boldsymbol{\lambda}^{c}_i \cdot
\boldsymbol{\lambda}^{c}_j
\left[\frac{1}{r_{ij}}-\frac{\pi}{2}\delta(\boldsymbol{r}_{ij})(\frac{1}{m^2_i}+\frac{1}{m^2_j}
+\frac{4\boldsymbol{\sigma}_i\cdot\boldsymbol{\sigma}_j}{3m_im_j})-\frac{3}{4m_im_jr^3_{ij}}
S_{ij}\right] \label{sala-vG} \\
V^{\chi}_{ij} & = & V_{\pi}( \boldsymbol{r}_{ij})\sum_{a=1}^3\lambda
_{i}^{a}\cdot \lambda
_{j}^{a}+V_{K}(\boldsymbol{r}_{ij})\sum_{a=4}^7\lambda
_{i}^{a}\cdot \lambda _{j}^{a}
+V_{\eta}(\boldsymbol{r}_{ij})\left[\left(\lambda _{i}^{8}\cdot
\lambda _{j}^{8}\right)\cos\theta_P-(\lambda _{i}^{0}\cdot
\lambda_{j}^{0}) \sin\theta_P\right] \label{sala-Vchi1} \\
V_{\chi}(\boldsymbol{r}_{ij}) & = & {\frac{g_{ch}^{2}}{{4\pi
}}}{\frac{m_{\chi}^{2}}{{\
12m_{i}m_{j}}}}{\frac{\Lambda _{\chi}^{2}}{{\Lambda _{\chi}^{2}-m_{\chi}^{2}}}}%
m_{\chi} \left\{(\boldsymbol{\sigma}_{i}\cdot
\boldsymbol{\sigma}_{j})
\left[ Y(m_{\chi}\,r_{ij})-{\frac{\Lambda_{\chi}^{3}}{m_{\chi}^{3}}}%
Y(\Lambda _{\chi}\,r_{ij})\right] \right.\nonumber \\
&& \left. +\left[H(m_{\chi}
r_{ij})-\frac{\Lambda_{\chi}^3}{m_{\chi}^3}
H(\Lambda_{\chi} r_{ij})\right] S_{ij} \right\}, ~~~~~~\chi=\pi, K, \eta, \\
S_{ij}&=&\left\{ 3\frac{(\boldsymbol{\sigma}_i
\cdot\boldsymbol{r}_{ij}) (\boldsymbol{\sigma}_j\cdot
\boldsymbol{r}_{ij})}{r_{ij}^2}-\boldsymbol{\sigma}_i \cdot
\boldsymbol{\sigma}_j\right\},\\
H(x)&=&(1+3/x+3/x^{2})Y(x),~~~~~~
 Y(x) =e^{-x}/x. \label{sala-vchi2}
\end{eqnarray}
\end{widetext}
Where $S_{ij}$ is quark tensor operator; $Y(x)$ and $H(x)$ are
standard Yukawa functions; $T_c$ is the kinetic
energy of the center of mass; $\alpha_{s}$ is the quark-gluon coupling constant; $g_{ch}$ is the coupling constant
for chiral field, which is determined from the $NN\pi$ coupling constant through
\begin{equation}
\frac{g_{ch}^{2}}{4\pi }=\left( \frac{3}{5}\right) ^{2}{\frac{g_{\pi NN}^{2}%
}{{4\pi }}}{\frac{m_{u,d}^{2}}{m_{N}^{2}}}\label{gch}.
\end{equation}
The other symbols in the above expressions have their usual meanings. Generally, we use the parameters from our
previous work of dibaryons~\cite{Gao,HuangHX3}. However, the model parameters used in the dibaryon calculation
can describe the ground baryons well, but cannot fit the masses of the ground mesons, especially the $K$ meson,
the obtained mass of which is much higher than the experimental value. This situation will lead to a consequence
that some bound states cannot decay to the open channels, because of the much larger mass of $K$.
To solve this problem, we adjust the quark-gluon coupling constant $\alpha_{s}$ of the $q\bar{q}$ pair, and
keep the other parameters unchanged. By doing this,
the parameters can describe the nucleon-nucleon and hyperon-nucleon interaction well, and at the same time,
it will lower the mass of $K$ to the experimental value. The model parameters are fixed by fitting the spectrum
of baryons and mesons we used in this work. The parameters of Hamiltonian are given in Table~\ref{parameters}.
Besides, a phenomenological color screening confinement potential is used here, and $\mu_{ij}$ is the color
screening parameter, which is determined by fitting the deuteron properties, $NN$ scattering phase shifts,
$N\Lambda$ and $N\Sigma$ scattering phase shifts, respectively, with
$\mu_{uu}=0.45$, $\mu_{us}=0.19$ and
$\mu_{ss}=0.08$, satisfying the relation,
$\mu_{us}^{2}=\mu_{uu}\mu_{ss}$~\cite{HuangHX3}. The calculated masses of baryons and
mesons in comparison with experimental values are shown in Table~\ref{mass}.

\begin{table}[ht]
\caption{\label{parameters}Model parameters:
$m_{\pi}=0.7$ fm$^{-1}$, $m_{ k}=2.51$ fm$^{-1}$,
$m_{\eta}=2.77$ fm$^{-1}$, $\Lambda_{\pi}=4.2$ fm$^{-1}$, $\Lambda_{K}=\Lambda_{\eta}=5.2$ fm$^{-1}$,
$g_{ch}^2/(4\pi)$=0.54, $\theta_p$=$-15^{0}$. }
\begin{tabular}{ccccccccc} \hline\hline
$b$ & ~~~$m_{u}$~~~~ & ~~~$m_{s}$~~~ & ~~~$a_c$~~~ & ~~~$V^{(qq)}_{0}$~~~ & ~~~$V^{(q\bar{q})}_{0}$~~~~   \\
($fm$) & ($MeV$) & ($MeV$) & ($MeV\cdot fm^{-2}$) & ($fm^{2}$) & ($fm^{2}$)   \\ \hline
0.518  & 313 & 573 & 58.03 &    -1.2883 & -0.2012  \\ \hline
$\alpha_{s}^{uu}$ &  $\alpha_{s}^{us}$ & $\alpha_{s}^{ss}$ & $\alpha_{s}^{u\bar{u}}$ &  $\alpha_{s}^{u\bar{s}}$ &
 $\alpha_{s}^{s\bar{s}}$   \\ \hline
  0.5652 & 0.5239 & 0.4506 & 1.7930 & 1.7829 & 1.5114 &\\ \hline \hline
\end{tabular}
\end{table}

\begin{table}[ht]
\caption{The masses (in MeV) of the baryons and mesons
obtained from QDCSM. Experimental values are taken
from the Particle Data Group (PDG)~\cite{PDG}.}
\begin{tabular}{lccccccccc}
\hline \hline
 & ~~$N$~~ & ~~$\Delta$~~ & ~~$\Lambda$~~& ~~$\Sigma$~~ & ~~$\Sigma^{*}$~~ & ~~$\Xi$~~ & ~~$\Xi^{*}$~~ & ~~$\Omega$~~  \\
\hline
 Expt.& 939 & 1232 & 1116 & 1193 & 1385 & 1318 & 1533 & 1672  \\
\hline
QDCSM & 939 & 1232 & 1124 & 1238 & 1360 & 1374 & 1496 & 1642  \\
\hline\hline
& ~~$\eta^{'}$~~ & ~~$K$~~ & ~~$K^{*}$~~ & ~~$\phi$~~    \\
\hline
 Expt.& 958 & 495 & 892 & 1020    \\
\hline
QDCSM & 852 & 495 & 892 & 1020   \\
\hline\hline
\end{tabular}
\label{mass}
\end{table}

The quark delocalization in QDCSM is realized by specifying the
single particle orbital wave function of QDCSM as a linear
combination of left and right Gaussians, the single particle
orbital wave functions used in the ordinary quark cluster model,
\begin{eqnarray}
\psi_{\alpha}(\mathbf{s}_i ,\epsilon) & = & \left(
\phi_{\alpha}(\mathbf{s}_i)
+ \epsilon \phi_{\alpha}(-\mathbf{s}_i)\right) /N(\epsilon), \nonumber \\
\psi_{\beta}(-\mathbf{s}_i ,\epsilon) & = &
\left(\phi_{\beta}(-\mathbf{s}_i)
+ \epsilon \phi_{\beta}(\mathbf{s}_i)\right) /N(\epsilon), \nonumber \\
N(\epsilon) & = & \sqrt{1+\epsilon^2+2\epsilon e^{-s_i^2/4b^2}}. \label{1q} \\
\phi_{\alpha}(\mathbf{s}_i) & = & \left( \frac{1}{\pi b^2}
\right)^{3/4}
   e^{-\frac{1}{2b^2} (\mathbf{r}_{\alpha} - \frac{2}{5}\mathbf{s}_i)^2} \nonumber \\
\phi_{\beta}(-\mathbf{s}_i) & = & \left( \frac{1}{\pi b^2}
\right)^{3/4}
   e^{-\frac{1}{2b^2} (\mathbf{r}_{\beta} + \frac{3}{5}\mathbf{s}_i)^2}. \nonumber
\end{eqnarray}
Here $\mathbf{s}_i$, $i=1,2,...,n$ are the generating coordinates,
which are introduced to expand the relative motion wavefunction~\cite{QDCSM0,QDCSM1}.
The mixing parameter $\epsilon(\mathbf{s}_i)$ is not an adjusted one but determined
variationally by the dynamics of the multi-quark system itself. In this way, the multi-quark
system chooses its favorable configuration in the interacting process. This mechanism has been
used to explain the cross-over transition between hadron phase and quark-gluon plasma phase~\cite{Xu}.

\section{The results and discussions}

In this work, we perform a dynamical investigation of the molecular states composed of $\Sigma$
(or $\Sigma^{*}$) and $K$ (or $K^{*}$) in the QDCSM. Our purpose is to understand the interaction
properties of the $\Sigma$ (or $\Sigma^{*}$) and $K$ (or $K^{*}$), and to see whether there
exists any $P_{c}$-like pentaquarks in hidden strange sector. Moreover, we also attempt to explore if there is any
pentaquark states which can be used to explain some nucleon resonances. For the system with isospin
$I=\frac{1}{2}$ and $J^{P}=\frac{1}{2}^{-}$, we investigate three molecular states $\Sigma K$,
$\Sigma K^{*}$ and $\Sigma^{*} K^{*}$; for the system with isospin $I=\frac{1}{2}$ and
$J^{P}=\frac{3}{2}^{-}$, we investigate three molecular states $\Sigma K^{*}$, $\Sigma^{*} K$ and $\Sigma^{*} K^{*}$.

Since an attractive potential is necessary for forming bound state or resonance, the effective
potentials between $\Sigma$ (or $\Sigma^{*}$) and $K$ (or $K^{*}$) are calculated and shown in Figs. 1.
The effective potential between two colorless clusters is defined as, $V(s)=E(s)-E(\infty)$, where
$E(s)$ is the diagonal matrix element of the Hamiltonian of the system in the generating coordinate.
For the $IJ^{P}=\frac{1}{2}\frac{1}{2}^{-}$ system (Fig. 1 (a)), one sees that the potentials
are all attractive for the channels $\Sigma K$, $\Sigma K^{*}$ and $\Sigma^{*} K^{*}$. The attraction between
$\Sigma^{*}$ and $K^{*}$ is the largest one, followed by that of the $\Sigma K^{*}$ channel, then the
$\Sigma K$ channel. This rule is very similar to the interactions between $\Sigma_c$ (or $\Sigma_c^{*}$) and $D$ (or $D^{*}$)~\cite{HuangHX1}. For the $IJ^{P}=\frac{1}{2}\frac{3}{2}^{-}$ system (Fig. 1 (b)), the
potentials are all attractive for channels $\Sigma K^{*}$, $\Sigma^{*} K$ and $\Sigma^{*} K^{*}$. The
attractions of both $\Sigma K^{*}$ and $\Sigma^{*} K^{*}$ channels are larger than that of the $\Sigma^{*} K$ channel.

\begin{center}
\begin{figure}
\epsfxsize=3.3in \epsfbox{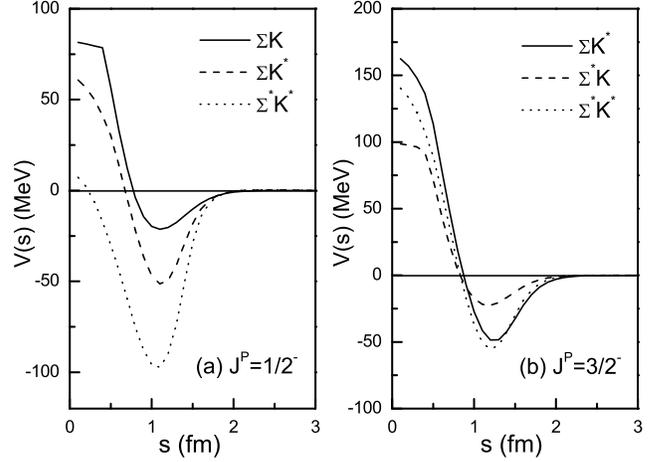} \vspace{-0.1in}

\caption{The potentials of different channels for the
$J^{P}=\frac{1}{2}\frac{1}{2}^{-}$ and $J^{P}=\frac{1}{2}\frac{3}{2}^{-}$ systems.}
\end{figure}
\end{center}

In order to see whether or not there is any bound state, a dynamic calculation is needed. The resonating group
method (RGM)~\cite{RGM}, a well established method for studying a bound-state problem or a scattering one,
is used here. The wave function of the baryon-meson system is of the form
\begin{equation}
\Psi = {\cal A } \left[\hat{\psi}_{A}(\boldsymbol{\xi}_{1},\boldsymbol{\xi}_{2})\hat{\psi}_{B}(\boldsymbol{\xi}_{3})\chi_{L}(\boldsymbol{R}_{AB})\right].
\end{equation}
where $\boldsymbol{\xi}_{1}$ and $\boldsymbol{\xi}_{2}$ are the internal coordinates for the baryon cluster A,
and $\boldsymbol{\xi}_{3}$ is the internal coordinate for the meson cluster B.
$\boldsymbol{R}_{AB} = \boldsymbol{R}_{A}-\boldsymbol{R}_{B}$ is the relative coordinate between the two
clusters A and B. The $\hat{\psi}_{A}$ and $\hat{\psi}_{B}$ are the antisymmetrized internal cluster wave
functions of the baryon A and meson B, and $\chi_{L}(\boldsymbol{R}_{AB})$ is the relative motion wave
function between two clusters. The symbol ${\cal A }$ is the anti-symmetrization operator defined as
\begin{equation}
{\cal A } = 1-P_{14}-P_{24}-P_{34}.
\end{equation}
where 1, 2, and 3 stand for the quarks in the baryon cluster and 4 stands for the quark in the meson cluster.
Here, we expand this relative motion wave function by gaussian bases
\begin{eqnarray}
\chi_{L}(\boldsymbol{R}_{AB}) &=& \frac{1}{\sqrt{4\pi}}(\frac{6}{5\pi b^2}) \sum_{i=1}^{n} C_{i}  \nonumber \\
&& \hspace{-1cm} \times  \int \exp\left[-\frac{3}{5b^2}(\boldsymbol{R}_{AB}-\boldsymbol{S}_{i})^{2}\right] Y_{LM}(\hat{\boldsymbol{S}_{i}})d\hat{\boldsymbol{S}_{i}}. \nonumber \\
\end{eqnarray}
where $\boldsymbol{S}_{i}$ is the generating coordinate, $n$ is the number of the gaussian bases,
which is determined by the stability of the results.
By doing this, the integro-differential equation of RGM can be reduced to algebraic equation, generalized
eigen-equation. Then we can obtain the energy of the system by solving this generalized eigen-equation.
The details of solving the RGM equation can be found in Ref.~\cite{RGM}. In our calculation, the distribution
of gaussians is fixed by the stability of the results. The results are stable when the largest distance between
the baryon-meson clusters is around 6 fm. To keep the dimensions
of matrix manageably small, the baryon-meson separation is taken to be less than 6 fm.

For the single channel calculations, the strong attractive interaction between $\Sigma$ (or $\Sigma^{*}$)
and $K$ (or $K^{*}$) leads to the total energy below the threshold of the two particles. All the binding energies
(labeled as $B$) and the masses (labeled as $M$) of molecular pentaquarks are listed in Table~\ref{bound}.
We need to mention that the mass of the bound state can be generally splitted into three terms: the baryon mass
$M_{baryon}$, the meson mass $M_{meson}$, and the binding energy $B$. To minimize the theoretical deviations,
the former two terms, $M_{baryon}$ and $M_{meson}$, are shifted to the experimental values.

\begin{table}
\begin{center}
\caption{The binding energy and masses (in MeV) of the molecular pentaquarks.}
{\begin{tabular}{@{}cc|cc} \hline
\multicolumn{2}{c|}{$J^{P}=\frac{1}{2}^{-}$}
 &\multicolumn{2}{c}{$J^{P}=\frac{3}{2}^{-}$}\\ \hline
 ~~~Channel~~~ & ~~~~$B/M$~~~~ & ~~~Channel~~~ & ~~~~$B/M$~~~~   \\
 {$\Sigma K$}      & $-18.8/1669.2$ &  {$\Sigma K^{*}$} & $-22.7/2062.3$    \\
 {$\Sigma K^{*}$}  & $-7.2/2077.8$  & {$\Sigma^{*}K$} & $-7.4/1872.6$  \\
 {~$\Sigma^{*} K^{*}$~}  & $-21.9/2255.1$  & {~$\Sigma^{*}K^{*}$~} & $-6.8/2270.2$   \\
  \hline
\end{tabular}
\label{bound}}
\end{center}
\end{table}

To confirm whether or not these bound states can survive as resonance states after coupling to the open channels, the study of the scattering process of the open channels is needed.
Resonances are unstable particles usually observed as bell-shaped structures in scattering cross sections of their open channels. For a simple narrow resonance, its fundamental properties correspond to the visible cross-section features: mass is the peak position, and decay width is the half-width of the bell shape. To find the resonance mass and decay width of the bound states showed in Table~\ref{bound}, we can calculate the cross-section of the corresponding open channels. The cross-section can be obtained from the scattering phase shifts by the formula:
\begin{equation}
\sigma = \frac{4\pi}{k^{2}}\cdot(2l+1)\cdot \sin^2\delta,  \label{sec}
\end{equation}
where $k=\sqrt{2\mu E_{cm}}/\hbar$; $\mu$ is the reduced mass of two hadrons of the open channel; $E_{cm}$ is the incident energy; $\delta$ is the scattering phase shift of the open channel, which can be obtained by the well-developed RGM~\cite{RGM}.

In this work, we study the pentaquarks composed of $udds\bar{s}$, so the open channels composed of $uddu\bar{u}$
are not considered at the present stage. For the $IJ^{P}=\frac{1}{2}\frac{1}{2}^{-}$ system, the bound state
$\Sigma K$ can be coupled to one open channel: the $S-$wave $\Lambda K$; the bound state $\Sigma K^{*}$ can be
coupled to eight open channels: the $S-$wave $N\eta^{\prime}$, $N\phi$, $\Lambda K$, $\Lambda K^{*}$, $\Sigma K$ and
the $D-$wave $N\phi$, $\Lambda K^{*}$, $\Sigma^{*} K$; the bound state $\Sigma^{*} K^{*}$ can be coupled to ten
open channels: the $S-$wave $N\eta^{\prime}$, $N\phi$, $\Lambda K$, $\Lambda K^{*}$, $\Sigma K$, $\Sigma K^{*}$ and
the $D-$wave $N\phi$, $\Lambda K^{*}$, $\Sigma K^{*}$, $\Sigma^{*} K$. All these open channels are listed in the first
column of Table~\ref{res1}, and the resonance states are listed in the first low of Table~\ref{res1}. We calculate
the scattering phase shifts of all these open channels, and then the cross-section by using the Eq.(\ref{sec}),
finally we can obtain the resonance mass and decay width of the resonance states, which are show in Table~\ref{res1}.
For the $IJ^{P}=\frac{1}{2}\frac{3}{2}^{-}$ system, we do the same calculation as that of the
$IJ^{P}=\frac{1}{2}\frac{1}{2}^{-}$ system, and all resonance states and the corresponding open channels,
as well as the resonance mass and decay width are shown in Table~\ref{res2}. To save space, here we only show the cross-section of all open channels for the state $\Sigma K^{*}$ with $J^{P}=\frac{3}{2}^{-}$ (see Fig. 2). The resonance mass and decay width of this state are obtained from the cross-section of those related open channels.
There are several features which are discussed below.

\begin{center}
\begin{figure}
\epsfxsize=3.3in \epsfbox{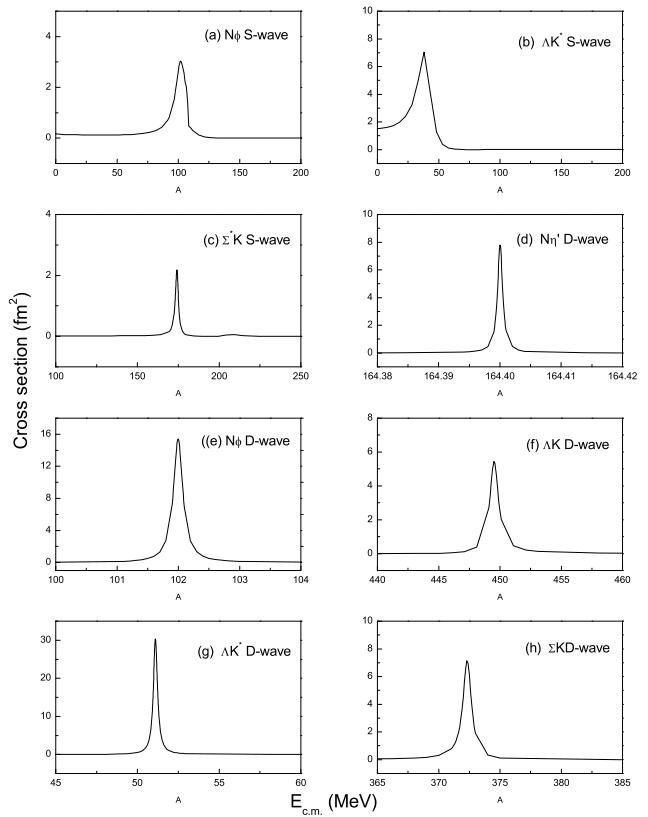} \vspace{-0.1in}

\caption{The cross-section of all open channels for the state $\Sigma K^{*}$ with $J^{P}=\frac{3}{2}^{-}$.}
\end{figure}
\end{center}

\begin{table}
\begin{center}
\caption{The resonance mass and decay width (in MeV) of the molecular pentaquarks with $J^{P}=\frac{1}{2}^{-}$.}
{\begin{tabular}{@{}c|cc|cc|cc} \hline
\multicolumn{1}{c|}{}
 &\multicolumn{2}{c|}{$\Sigma K$}&\multicolumn{2}{c|}{$\Sigma K^{*}$}&\multicolumn{2}{c}{$\Sigma^{*} K^{*}$}\\ \hline
 ~~~{$S-$wave}~~~ & ~~~$M_{r}$~~~ & ~~~$\Gamma_{i}$~~~ & ~~~$M_{r}$~~~ & ~~~$\Gamma_{i}$~~~ & ~~~$M_{r}$~~~ & ~~~$\Gamma_{i}$~~~   \\
 {$N \eta^{\prime}$}   & $-$ & $-$  & $2079.4$ & $1.1$ & $2246.8$ & $20.0$    \\
 {$N \phi$}       & $-$ & $-$  & $2080.0$ & $3.6$ & $2237.0$ & $30.0$    \\
 {$\Lambda K$}    & $1668.0$ & $1.3$  & $2083.4$ & $1.0$ & $2261.5$ & $20.0$    \\
 {$\Lambda K^{*}$}& $-$ & $-$  & $2056.6$ & $0.2$ & $2219.0$ & $58.0$    \\
 {$\Sigma K$}     & $-$ & $-$  & $2071.6$ & $4.6$ & $2252.3$ & $6.0$    \\
 {$\Sigma K^{*}$} & $-$ & $-$  & $-$      & $-$   & $2253.9$ & $16.0$    \\
 {$D-$wave} &  &   &  &  &  &    \\
 {$N \phi$}       & $-$ & $-$  & $2076.3$ & $0.3$ & $2254.4$ & $0.006$    \\
 {$\Lambda K^{*}$}& $-$ & $-$  & $2076.3$ & $0.4$ & $2253.6$ & $0.6$    \\
 {$\Sigma K^{*}$} & $-$ & $-$  & $-$      & $-$   & $2254.0$ & $0.06$    \\
 {$\Sigma^{*} K$} & $-$ & $-$  & $2076.8$ & $0.01$ & $2253.3$ & $0.8$    \\
  \hline
\end{tabular}
\label{res1}}
\end{center}
\end{table}

\begin{table}
\begin{center}
\caption{The resonance mass and decay width (in MeV) of the molecular pentaquarks with $J^{P}=\frac{3}{2}^{-}$.}
{\begin{tabular}{@{}c|cc|cc|cc} \hline
\multicolumn{1}{c|}{}
 &\multicolumn{2}{c|}{$\Sigma K^{*}$}&\multicolumn{2}{c|}{$\Sigma^{*} K$}&\multicolumn{2}{c}{$\Sigma^{*} K^{*}$}\\ \hline
 ~~~{$S-$wave}~~~ & ~~~$M_{r}$~~~ & ~~~$\Gamma_{i}$~~~ & ~~~$M_{r}$~~~ & ~~~$\Gamma_{i}$~~~ & ~~~$M_{r}$~~~ & ~~~$\Gamma_{i}$~~~   \\
 {$N \phi$}       & $2060.6$ & $10.4$  & $-$ & $-$ & $2270.5$ & $0.03$    \\
 {$\Lambda K^{*}$}& $2046.1$ & $15.0$  & $-$ & $-$ & $2256.5$ & $2.0$    \\
 {$\Sigma K^{*}$} & $-$ & $-$  & $-$   & $-$   & $2270.6$ & $0.1$    \\
 {$\Sigma^{*} K$} & $2054.1$ & $2.3$  & $-$ & $-$ & $2263.6$ & $3.7$    \\
 {$D-$wave} &  &   &  &  &  &    \\
 {$N \eta^{\prime}$}   & $2061.4$ & $0.001$  & $1875.7$ & $0.0004$ & $2269.2$ & $0.01$    \\
 {$N \phi$}       & $2061.0$ & $0.2$  & $-$ & $-$ & $2269.3$ & $0.01$    \\
 {$\Lambda K$}    & $2060.6$ & $0.9$  & $1871.6$ & $0.08$ & $2269.2$ & $0.02$    \\
 {$\Lambda K^{*}$}& $2059.1$ & $0.3$  & $-$ & $-$ & $2269.1$ & $0.05$    \\
 {$\Sigma K$}     & $2060.3$ & $0.9$  & $1871.6$ & $0.05$ & $2269.2$ & $0.02$    \\
 {$\Sigma K^{*}$} & $-$ & $-$  & $-$      & $-$   & $2269.2$ & $0.003$    \\

  \hline
\end{tabular}
\label{res2}}
\end{center}
\end{table}

First, the bound states showed in Table~\ref{bound} are all resonance states by coupling the corresponding
open channels. Because only the hidden strange channels are considered here, the total decay width of the states
given below is the lower limits. For the $IJ^{P}=\frac{1}{2}\frac{1}{2}^{-}$ system, the resonance mass of
$\Sigma K$ is $1668.0$ MeV, and the decay width is very small which is only $1.3$ MeV; the $\Sigma K^{*}$ is also
possible a narrow resonance state with the mass range of $2056.6 \sim 2083.4$ MeV and the decay width is $\sim 10$ MeV;
the mass of the resonance $\Sigma^{*} K^{*}$ is between $2219.0 \sim 2261.5$ MeV, while the decay width is much
larger, which is about $150$ MeV at least. For the $IJ^{P}=\frac{1}{2}\frac{3}{2}^{-}$ system, both the
$\Sigma^{*} K$ and $\Sigma^{*} K^{*}$ are very narrow resonance states with the mass range of $1871.6 \sim 1875.7$ MeV
and $2256.5 \sim 2270.5$ MeV respectively. Besides, the resonance mass range of $\Sigma K^{*}$ state is $2046.1
\sim 2061.4$ MeV and the decay width is about $30$ MeV.

Secondly, it is obvious that the decay width of decaying to $D-$wave channels is much smaller than that of
decaying to the $S-$wave channels. This is reasonable. In our quark model calculation, the coupling between
$S-$wave channels is through the central force, while the coupling between $S-$ and $D-$wave channels is
dominated by the tensor force, and the effect of the tensor force is much smaller than that of the central force.
This conclusion is consistent with our previous calculation of the dibaryon systems~\cite{HuangHX4,HuangHX5}.
Besides, we only consider the two-body decay channels in this work. The calculation of more decay channels will
change the total decay width of the resonance states.

Thirdly, our results in the hidden strange sector is similar to our previous study of the hidden charm molecular pentaquarks~\cite{HuangHX1}. In Ref.~\cite{HuangHX1}, we found that three states with $J^{P}=\frac{1}{2}^{-}$:
$\Sigma_{c}D$, $\Sigma_{c}D^{*}$, and $\Sigma^{*}_{c}D^{*}$ , and the other three states with $J^{P}=\frac{3}{2}^{-}$:
$\Sigma_{c}D^{*}$, $\Sigma^{*}_{c}D$, and $\Sigma^{*}_{c}D^{*}$  were all quasi-stable states. Analogously,
in this work, we find that three states with $J^{P}=\frac{1}{2}^{-}$: $\Sigma K$, $\Sigma K^{*}$, and $\Sigma^{*}K^{*}$ ,
and the other three states with $J^{P}=\frac{3}{2}^{-}$ : $\Sigma K^{*}$, $\Sigma^{*} K$, and $\Sigma^{*} K^{*}$ are
all resonance states. Besides, in Ref.~\cite{HuangHX1}, the molecular pentaquark $\Sigma^{*}_{c}D$ with quantum numbers
$IJ^{P}=\frac{1}{2}\frac{3}{2}^{-}$ can be used to explain the LHCb $P_{c}(4380)$ state. So here, the molecular
pentaquark $\Sigma^{*} K$ with quantum numbers $IJ^{P}=\frac{1}{2}\frac{3}{2}^{-}$ can be seen as a strange partner of
the LHCb $P_{c}(4380)$ state. This conclusion is consistent with the work on the hadron level~\cite{HeJ5}.

Finally, we find the mass of the $\Sigma^{*} K$ with quantum numbers $IJ^{P}=\frac{1}{2}\frac{3}{2}^{-}$ is close to the nucleon resonance $N^{*}(1875)$ listed in PDG~\cite{PDG}. Obviously, both the mass and the quantum numbers of this molecular pentaquark $\Sigma^{*} K$ correspond to the $N^{*}(1875)$. This conclusion is also consistent with the work on the hadron level~\cite{HeJ5}, in which the molecular state $\Sigma^{*} K$ was investigated in a quasipotential Bethe-Saltpeter equation approach and it was identified as the $N^{*}(1875)$ listed in PDG. Moreover, Ref.~\cite{HeJ5} also found that the $\Sigma K^{*}$ interaction produced a bound state with quantum numbers $IJ^{P}=\frac{1}{2}\frac{3}{2}^{-}$, which was related to the experimentally observed $N^{*}(2100)$ in the $\phi$ photo-production. Our results of the $\Sigma K^{*}$ state in the quark level is also consistent with that of Ref.~\cite{HeJ5}, so we also support that the molecular pentaquark $\Sigma K^{*}$ can be identified as the $N^{*}(2100)$.

\section{Summary}

In summary, we perform a dynamical investigation of the molecular states composed of $\Sigma$ (or $\Sigma^{*}$) and
$K$ (or $K^{*}$) within the QDCSM. We calculate the effective potential, the mass and decay widths of these molecular
states. Our results show: (1) The interactions between $\Sigma$ (or $\Sigma^{*}$) and $K$ (or $K^{*}$) are strong
enough to form the bound states, which are $\Sigma K$, $\Sigma K^{*}$ and $\Sigma^{*} K^{*}$ with
$IJ^{P}=\frac{1}{2}\frac{1}{2}^{-}$ and $\Sigma K^{*}$, $\Sigma^{*} K$ and $\Sigma^{*} K^{*}$ with
$IJ^{P}=\frac{1}{2}\frac{3}{2}^{-}$. And all these states are transferred to the resonance states by coupling the
open channels. (2) Our results in the hidden strange sector is similar to our previous study of the hidden charm
molecular pentaquarks~\cite{HuangHX1}, and the molecular pentaquark $\Sigma^{*} K$ with quantum numbers $IJ^{P}=\frac{1}{2}\frac{3}{2}^{-}$ can be seen as a strange partner of the LHCb $P_{c}(4380)$ state.
(3) This $\Sigma^{*} K$ state can also be identified as the nucleon resonance $N^{*}(1875)$ listed in PDG.
The $\Sigma K^{*}$ with quantum numbers $IJ^{P}=\frac{1}{2}\frac{3}{2}^{-}$ can be identified as the
$N^{*}(2100)$, which was experimentally observed in the $\phi$ photo-production.

In this work, we only study the pentaquarks composed of $udds\bar{s}$, so the open channels composed of $uddu\bar{u}$ are not considered at the present stage. Besides, we only consider the two-body decay channels. The calculation of more decay channels will change the total decay width of the resonance states. We will do this work in future.

\section*{Acknowledgment}

This work is supported partly by the National Science Foundation
of China under Contract Nos. 11675080, 11775118 and 11535005, the Natural Science Foundation of
the Jiangsu Higher Education Institutions of China (Grant No. 16KJB140006).


\begin{thebibliography}{99}
\bibitem{LHCb} R. Aaij, {\em et al.} (LHCb Collaboration), Phys. Rev. Lett. {\bf 115}, 072001 (2015).
\bibitem{Jlab} Z. E. Meziani, {\em et al.}, arXiv:1609.00676v2.
\bibitem{ChenHX} H. X. Chen, W. Chen, X. Liu and S. L. Zhu, Phys. Rep. {\bf 639}, 1 (2016).
\bibitem{ChenR} R. Chen, X. Liu, X. Q. Li and S. L. Zhu, Phys. Rev. Lett. {\bf 115}, 132001 (2015).
\bibitem{HeJ1} J. He, Phys. Lett. B {\bf 753}, 547 (2016).
\bibitem{ChenHX2} H. X. Chen, W. Chen, X. Liu, T. G. Steel and S. L. Zhu, Phys. Rev. Lett. {\bf 115}, 172001 (2015).
\bibitem{Roca} L. Roca, J. Nieves and E. Oset, Phys. Rev. D {\bf 92}, 094003 (2015).
\bibitem{Oller} Ulf-G. Mei{\ss}ner and J. A. Oller, Phys. Lett. B {\bf 751}, 59 (2015).
\bibitem{Kubarovsky} V. Kubarovsky and M. B. Voloshin, Phys. Rev. D {\bf 92}, 031502 (2015).
\bibitem{Guo} F. K. Guo, Ulf-G. Mei{\ss}ner, W. Wang and Z. Yang, Phys. Rev. D {\bf
92}, 071502 (2015).
\bibitem{Liuxh} X. H. Liu, Q. Wang and Q. Zhao, Phys. Lett. B {\bf 757}, 231 (2016).
\bibitem{Burns} T. J. Burns, Eur. Phys. J. A {\bf 51}, 152 (2015).
\bibitem{HuangHX1} H. X. Huang, C. R. Deng, J. L. Ping and F.
Wang, Eur. Phys. J. C {\bf 76}, 624 (2016).
\bibitem{Gao} H. Gao, T. -S. H. Lee, and V. Marinov, Phys. Rev. C {\bf 63}, 022201(R) (2001).
\bibitem{Brodsky} S. J. Brodsky, I. Schmidt and G. F. de Teramond, Phys. Rev. Lett. {\bf 64}, 1011 (1990).
\bibitem{Luke} M. E. Luke, A. V. Manohar, and M. J. Savage, Phys. Lett. B {\bf 288}, 355 (1992).
\bibitem{Liska} S. Liska, H. Gao, W. Chen, and X. Qian, Phys. Rev. C {\bf 75}, 058201 (2007).
\bibitem{Chiral} F. Huang, Z. Y. Zhang, and Y. W. Yu, Phys. Rev. C {\bf 73}, 025207 (2006).
\bibitem{Lattic} S. R. Beane, E. Chang, S. D. Cohen, W. Detmold, H.-W. Lin, K. Orginos, A. Parreno and M. J. Savage, Phys. Rev. D {\bf 91}, 114503 (2015).
\bibitem{Xie} J. J. Xie and F. K. Guo, Phys. Lett. B {\bf 774}, 108 (2017).
\bibitem{Gao2} H. Gao, H. Huang, T. Liu, J. Ping, F. Wang and Z. Zhao, Phys. Rev. C {\bf 95}, 055202 (2017).
\bibitem{Dalitz} R. H. Dalitz and S. F. Tuan, Ann. Phys. (N.Y.) {\bf 10}, 307 (1960).
\bibitem{Kaiser} N. Kaiser, P. B. Siegel, and W. Weise, Nucl. Phys. A {\bf 594}, 325 (1995).
\bibitem{Oset} E. Oset, and A. Ramos, Nucl. Phys. A {\bf 635}, 99 (1998).
\bibitem{Krippa} B. Krippa, Phys. Rev. C {\bf 58}, 1333 (1998).
\bibitem{Oller2} J. A. Oller and Ulf-G. Mei{\ss}ner, Phys. Lett. B {\bf 500}, 263 (2001).
\bibitem{Jido} D.Jido. T. Sekihara, Y. Ikeda, T. Hyodo, Y. Kanada-En'yo, and E. Oset, Nucl. Phys. A {\bf 835}, 59 (2010).
\bibitem{Hall} J. M. M. Hall, W. Kamleh, D. B. Leinweber, B. J. Menadue, B. J. Owen, A. W. Thomas, and R. D. Young, Phys. Rev. Lett. {\bf 114}, 132002 (2015).
\bibitem{Inoue} T. Inoue, E. Oset, and M.J. Vicente Vacas, Phys. Rev. C {\bf 65}, 035204 (2002).
\bibitem{LiuBC} B. C. Liu and B. S. Zou, Phys. Rev. Lett. {\bf 96}, 042002 (2006).
\bibitem{Helminen} C. Helminen and D.O. Riska, Nucl. Phys. A {\bf 699}, 624 (2002).
\bibitem{ZouBS1} B. S. Zou, Eur. Phys. J. A {\bf 35}, 325 (2008).
\bibitem{AnCS} C. S. An and B. S. Zou, Eur. Phys. J. A {\bf 39}, 195 (2009).
\bibitem{ZouBS2} B. S. Zou, Nucl. Phys. A {\bf 835}, 199 (2010).
\bibitem{PDG} C. Patrignani, {\em et al.}, Particle Data Group, Chin. Phys. C {\bf 40}, 100001 (2016).
\bibitem{Kashevarov} V. L. Kashevarov, et al., Phys. Rev. Lett. {\bf 118}, 212001 (2017).
\bibitem{Collins} P. Collins, {\em et al.}, Phys. Lett. B {\bf 771}, 213 (2017).
\bibitem{Anisovich} A. V. Anisovich, R. Beck, E. Klempt, V. A. Nikonov, A. V. Sarantsev, and U. Thoma, Eur. Phys. J. A {\bf 48}, 15 (2012).
\bibitem{HeJ2} J. He, Phys. Rev. C {\bf 86}, 035204 (2012).
\bibitem{HeJ3} J. He, Phys. Rev. C {\bf 91}, 018201 (2015).
\bibitem{HeJ4} J. He, Nucl. Phys. A {\bf 927}, 24 (2014).
\bibitem{Mibe} T. Mibe, {\em et al.} (LEPS Collaboration), Phys. Rev. Lett. {\bf 95}, 182001 (2005).
\bibitem{Kiswandhi1} A. Kiswandhi, J. J. Xie, and S. N. Yang, Phys. Lett. B {\bf 691}, 214 (2010).
\bibitem{Kiswandhi2} A. Kiswandhi and S. N. Yang, Phys. Rev. C {\bf 86}, 015203 (2012).
\bibitem{Chang1} W. C. Chang, {\em et al.} (LEPS Collaboration), Phys. Rev. C {\bf 82}, 015202 (2010).
\bibitem{Chang2} W. C. Chang, {\em et al.} (LEPS Collaboration), Phys. Lett. B {\bf 684}, 6 (2010).
\bibitem{Qian} X. Qian, {\em et al.}, Phys. Lett. B {\bf 696}, 338 (2011).
\bibitem{Kiswandhi3} A. Kiswandhi, S. N. Yang, and Y. B. Dong, Phys. Rev. C {\bf 94}, 015202 (2016).
\bibitem{HeJ5} J. He, Phys. Rev. D {\bf 95}, 074031 (2017).
\bibitem{RGM} M. Kamimura, Supp. Prog. Theo. Phys. {\bf 62}, 236 (1977).
\bibitem{Ping2009} J. L. Ping, H. X. Huang, H. R. Pang, F. Wang and C. W. Wong, Phys. Rev.
C {\bf 79}, 024001 (2009).
\bibitem{QDCSM0} F. Wang, G. H. Wu, L. J. Teng and T. Goldman, Phys. Rev. Lett. {\bf 69}, 2901 (1992);
\bibitem{QDCSM1} J. L. Ping, F. Wang and T. Goldman, Nucl. Phys. A {\bf 657}, 95 (1999); G. H. Wu, J. L. Ping,
L. J. Teng {\em et al.}, Nucl. Phys. A {\bf 673}, 279 (2000); H.
R. Pang, J. L. Ping, F. Wang and T. Goldman, Phys. Rev. C {\bf
65}, 014003 (2001); J. L. Ping, F. Wang and T. Goldman, Nucl.
Phys. A {\bf 688}, 871 (2001); J. L. Ping, H. R. Pang, F. Wang and
T. Goldman, Phys. Rev. C {\bf 65}, 044003 (2002).
\bibitem{HuangHX2} H. X. Huang, P. Xu, J. L. Ping and F.
Wang, Phys. Rev. C {\bf 84}, 064001 (2011).
\bibitem{HuangHX3} H. X. Huang, J. L. Ping and F. Wang, Phys. Rev. C {\bf 92}, 065202 (2015).
\bibitem{Xu} M. M. Xu, M. Yu and L. S. Liu, Phys. Rev. Lett. {\bf 100},
  092301 (2008).
\bibitem{HuangHX4} H. X. Huang, J. L. Ping and F. Wang, Phys. Rev. C {\bf 87}, 034002 (2013).
\bibitem{HuangHX5} H. X. Huang, J. L. Ping and F. Wang, Phys. Rev. C {\bf 89}, 035201 (2014).
\end{thebibliography}
\end{document}